\documentclass[12pt]{article}
\usepackage{ametsoc}
\usepackage{epstopdf}

\newcommand{\myabstract}{
Precipitation generates small-scale turbulent air flows the energy of which ultimately dissipates to heat. The power of
this process has previously been estimated to be around 2-4 W~m$^{-2}$ in the tropics: a value comparable in magnitude to
the dynamic power of the global circulation.
Here we suggest that this previous power estimate is approximately double the true figure.
Our result reflects a revised evaluation of the mean precipitation path length $H_P$. We investigate the dependence of $H_P$ on surface temperature,
relative humidity, temperature lapse rate and degree of condensation in the ascending air.
We find that the degree of condensation, defined as the relative change of the saturated water vapor mixing ratio in the region of condensation,
is a major factor determining $H_P$.
We estimate from theory that the mean large-scale rate of frictional dissipation associated with total precipitation in the tropics lies between 1 and 2 W~m$^{-2}$
and show that our estimate is supported by empirical evidence.
We show that under terrestrial conditions frictional dissipation constitutes a minor fraction of the dynamic power of condensation-induced atmospheric circulation,
which is estimated to be at least 2.5 times larger. However, because $H_P$ increases with surface temperature $T_s$,
the rate of frictional dissipation would exceed that of condensation-induced dynamics, and thus block major circulation, at $T_s \gtrsim 320$~K
in a moist adiabatic atmosphere.
}
\begin{document}
%
%
\title{\textbf{\large{The key physical parameters governing frictional dissipation in a precipitating atmosphere}}}
%
%
\author{\textsc{A. M. Makarieva, V. G. Gorshkov, A. V. Nefiodov}\\
\textit{\footnotesize{Theoretical Physics Division, Petersburg Nuclear Physics Institute, 188300, Gatchina,
St. Petersburg, Russia}}
\and
\centerline{\textsc{D. Sheil}\thanks{\textit{Corresponding author.}
                \newline{E-mail: D.Sheil@cgiar.org}}}\\
\centerline{\textit{\footnotesize{School of Environmental Science and Management, Southern Cross University, PO Box 157, Lismore, NSW 2480, Australia}}}\\
\centerline{\textit{\footnotesize{Institute of Tropical Forest Conservation, Mbarara University of Science and Technology, PO Box, 44, Kabale, Uganda}}}\\
\centerline{\textit{\footnotesize{Center for International Forestry Research, PO Box 0113 BOCBD, Bogor 16000, Indonesia}}}
\and
\centerline{\textsc{A. D. Nobre}}\\
\centerline{\textit{\footnotesize{Centro de Ci\^{e}ncia do Sistema Terrestre INPE, S\~{a}o Jos\'{e} dos Campos SP 12227-010, Brazil}}}
\and
\centerline{\textsc{P. Bunyard}}\\
\centerline{\textit{\footnotesize{Lawellen Farm, Withiel, Bodmin, Cornwall, PL30 5NW, United Kingdom and University Sergio Arboleda, Bogota, Colombia}}}
\and
\centerline{\textsc{B.-L. Li}}\\
\centerline{\textit{\footnotesize{XIEG-UCR International Center for Arid Land Ecology, University of California, Riverside, USA}}}
}
%
\ifthenelse{\boolean{dc}}
{
\twocolumn[
\begin{@twocolumnfalse}
\amstitle

\begin{center}
\begin{minipage}{13.0cm}
\begin{abstract}
    \myabstract
    \newline
    \begin{center}
        \rule{38mm}{0.2mm}
    \end{center}
\end{abstract}
\end{minipage}
\end{center}
\end{@twocolumnfalse}
]
}
{
\amstitle
\begin{abstract}
\myabstract
\end{abstract}
\newpage
}
\section{Introduction}
\label{sec1}
Understanding the physics of a moist atmosphere and capturing it in theoretical concepts is a major challenge for climate science \citep{sc06,sc10,co11}. Among the complications introduced by water vapour are the various influences of precipitation on atmospheric motion.  One specific aspect is that precipitation generates small-scale air turbulence around the falling condensate particles. The energy for this turbulent air motion derives from the potential energy of the rain drops or ice particles (i.e. "hydrometeors") in the gravitational field of Earth and is ultimately dissipated to heat. If this energy were not converted to turbulent kinetic energy of the air, the hydrometeors would continue to accelerate as they fall. But air exerts a drag force that prevents this acceleration. This force grows with increasing size of the hydrometeor and its velocity $W$ relative to the surrounding air. Thus, as the condensate particle is accelerated by gravity, this opposing force grows until it equals the weight of the particle. Acceleration ceases at this resulting terminal velocity $W_t$. Since hydrometeors fall at near their terminal velocities for most of the duration of their falls, the mean drag force acting on them over this period is approximately equal to their weight.

Consider a column of moist air as a mixture of dry air and water vapor which we denote here using subscripts $d$ and $v$ respectively,
standing for dry air and water vapor. In hydrostatic equilibrium,
\begin{equation}
\label{he}
-\frac{\partial p}{\partial z} = \rho g,
\end{equation}
where $p = p_d + p_v$ is air pressure, $\rho = \rho_d + \rho_v$ is air density, $g$ is the acceleration of gravity.
There is no available potential energy in such a column. We now cool this column in such a manner that some water vapor condenses.
Two types of potential energy have now become available: the potential energy of droplets in the gravitational field and the potential energy of any non-equilibrium air pressure gradient that may have formed upon condensation. We recently proposed that the release of the second type of potential energy, i.e. that associated with the non-equilibrium gradient of saturated water vapor, is a major driver of atmospheric circulation on Earth \citep{mg07,mg10,g12}. Since precipitation and, hence, frictional dissipation of the potential energy of hydrometeors, always accompany condensation, it is important to estimate and contrast the power of the two processes in order to understand their relative influences.

In this paper we first examine how the power $D$ of precipitation-related frictional dissipation can be estimated from basic atmospheric parameters.
We then compare our results to those of \citet{pa00} and explain why our estimates are more consistent with both theory and data.
We also discuss why the recent estimate of $D$ by \citet{pa12} from satellite-derived tropical rain rates does not constrain the true value more accurately than our theoretical analysis. We show that $D$ grows with increasing surface temperature and estimate the critical temperature when the power of frictional dissipation equals the power of condensation-induced dynamics, such that the latter ceases.

\section{Basic formulae}
\label{sec2}

The power of dissipation of energy $D$ (W~m$^{-2}$) per unit area associated with small-scale turbulence around hydrometeors can be written as
\begin{equation}
\label{D}
D = \int_0^\infty W F_c dz = \int_0^\infty W_t g \rho_c dz.
\end{equation}
Here $W \equiv w_c - w > 0$
is the mean velocity of hydrometeors relative to the air, $w_c$ and $w$ are the vertical velocities of
condensate particles and air relative to the Earth's surface, $F_c$ is the turbulent drag force per unit air volume exerted by air on hydrometeors, $\rho_c = N_c m$ is condensate density, $N_c$ is the number of hydrometeors per unit volume and $m$ is their mean mass. The second equality in (\ref{D}) takes into account our assumption that hydrometeors are falling at their terminal velocity $W_t$, such that the drag force acting on a droplet is equal to its weight. For a constant $W_t$ Eq.~(\ref{D}) simply represents the product of $W_t g$ and the total amount of condensate in the atmospheric column.

Equation (\ref{D}) is not suited for a theoretical analysis. The distribution of $W_t$, which depends strongly on particle size, is poorly known. The amount of condensate in the atmosphere varies greatly in time and space, from a few kilograms per square meter in severe storms to less than a hundred grams per square meter under normal conditions \citep[e.g.,][]{ji08,wo02}. Nonetheless, as we shall now show, it is possible to modify Eq.~(\ref{D}) to exclude these uncertain parameters.

Let us first consider the case when the terminal velocity of hydrometeors is much larger than air velocity, $W_t \gg |w|$. Neglecting for now re-evaporation of condensate in the column we assume that all condensate that has formed in the course of the ascent of moist air precipitates to the ground. In such a case the power $D$ becomes
\begin{equation}
\label{Dg}
D = \rho_{cs} W_{ts} g H_P = P_s g H_P.
\end{equation}
Here $\rho_{cs}$ and $W_{ts}$ are respectively the condensate density and the terminal velocity of hydrometeors near the ground surface, precipitation path length $H_P$ is the mean height from which the hydrometeors are falling, $P_s = \rho_{cs} W_{ts}$ is precipitation measured at the surface in kg~H$_2$O~m$^{-2}$~s$^{-1}$. When a hydrometeor falls to the ground from height $H_P$, the potential energy that it has lost per unit mass is $g H_P$. Precipitation $P_s$ tells us how much water hits the ground per unit surface area per unit time, so $D = P_s g H_P$ gives the total rate of potential energy loss by all hydrometeors in the column.

At $W_t \gg |w|$ there is no upward transport of condensate that originated in the lower atmospheric layers: the hydrometeors fall to the ground from where they were formed. Rate of condensation $S$ (mol~H$_2$O~m$^{-3}$~s$^{-1}$) in the ascending air is equal to
\begin{equation}
\label{S}
S = -w\Big(\frac{\partial N_v}{\partial z} - \gamma \frac{\partial N}{\partial z}\Big) \equiv -wN\frac{\partial \gamma}{\partial z} > 0,
\quad \gamma \equiv \frac{N_v}{N} = \frac{p_v}{p},
\end{equation}
where $N_v$ and $N$ are the molar densities of water vapor and air \citep{mg10,g12}. The value of $S$ differs from zero only in a certain area $z_1 \leqslant z \leqslant z_2$, where the relative humidity is close to unity and the water vapor is saturated. Precipitation path length $H_P$ is then equal to the mean height of condensation:
\begin{equation}
\label{HP}
H_P = \frac{\int\limits_0^\infty S(z) z dz}{\int\limits_0^\infty S(z) dz} =
\frac{\gamma^*(z_1)z_1 - \gamma^*(z_2)z_2+ \int\limits_{z_1}^{z_2} \gamma^*(z) dz}{\gamma^*(z_1)-\gamma^*(z_2)}.
\end{equation}
Here $z_1$ and $z_2$ are the heights of the lower and upper boundaries of the condensation area ($S = 0$ at $z <z_1$ and $z > z_2$), $\gamma = \gamma^* \equiv p_v^*/p$ is equal to the ratio of saturated water vapor partial pressure $p_v^*$ to air pressure $p$. (Note that $\gamma^* = (M_d/M_v) q^*/[1+q^* (M_d/M_v)] \simeq (M_d/M_v) q^*$, where $q^* \equiv \rho_v^*/\rho_d$ is the saturated water vapor mixing ratio, $M_v$ and $M_d$ are the molar masses of the water vapor and dry air, respectively. Thus, replacing $\gamma^*$ by $q^*$ in (\ref{HP}) will not significantly affect the estimate of $H_P$ at $q^* \ll 1$.) Condensation rate $S$ normalized by the integral in the denominator of (\ref{HP}) is the probability that a given hydrometeor reaching the surface has fallen from a height between $z$ and $z + dz$. The last expression in (\ref{HP}) is obtained by integrating the first expression by parts and taking into account that the upward flux of air $w N$ is approximately independent of $z$, that is, $\partial (Nw)/\partial z = 0$ (see Appendix for details).

We note that Eqs.~(\ref{S}) and (\ref{HP}) assume that $\gamma$ decreases with height solely because of condensation that occurs in the rising air parcel; possible mixing with ambient air that can lead to a change in $\gamma$ is neglected. In the real atmosphere the integration in (\ref{HP})
should be made over those height intervals where the mean ambient relative humidity is sufficiently high for condensation to occur,  e.g., a threshold
of $R_H(z) > 80\%$ can be applied \citep{lo90}.
Under the assumption of constant $Nw$, the vertical profile of precipitation $P(z) = \int_z^{z_2}S(z')dz'$ coincides
in form with the vertical profile of $\gamma^*(z)$ in the area of condensation: we should observe $P(z)/P_s = \gamma^*(z)/\gamma^*(z_1)$ at
$z_1 \leqslant z \leqslant z_2$.
Where the relative humidity is low but re-evaporation can be neglected, we should observe $P(z)/P_s$ stays approximately constant
independent of changes in $\gamma(z)$.

\section{Numerical estimates of $H_P$}
\label{sec3}

The advantage of Eq.~(\ref{HP}) is that $H_P$ can be estimated from theory.
Water vapor condenses as the moist saturated air ascends and cools.
We allow for the incompleteness of condensation which we can define as
$\zeta \equiv \gamma^*(z_2)/\gamma^*(z_1)$, which describes
the share of water vapor that has not condensed when the saturated air parcel rising from $z_1$ has reached $z_2$.
 For a fully saturated atmosphere with surface relative humidity of $R_H=100\%$ we have $z_1 = 0$, $z_2 = \infty$ in (\ref{HP}) and $\zeta = 0$.
When the vertical distribution of water vapor follows the moist adiabat, $H_P$ is unambiguously determined by the surface temperature (see Appendix). In Fig.~\ref{fig1}A we show the dependence of moist adiabatic $H_P$ on surface temperature $T_s$ at a relative humidity of $R_H = 100\%$ and three values of $\zeta$: 0, 1/2 and 2/3.

We can see that $H_P$ grows with increasing surface temperature $T_s$, e.g. at $\zeta = 0$ we have $H_P = 3.5$~km at 290~K and 5.3~km at 300~K, i.e. $H_P$ increases
by around 50\% for a 10 degrees rise in surface temperature. We also find that $H_P$ decreases sharply with increasing $\zeta$: at 300~K and $\zeta = 1/2$ we have $H_P = 2.4$~km, Fig.~\ref{fig1}A, i.e. $H_P$ decreases more than two-fold compared to the case of complete condensation $\zeta = 0$. Height $z_2$ at $\zeta = 1/2$ and $T_s = 300$~K is equal to 4.9~km, Fig.~\ref{fig1}A.

In the real atmosphere the lower layer $z < z_1$ is generally undersaturated with a global mean relative humidity at the surface of about $R_H =80\%$.
Value of $z_1$ depends on the temperature lapse rate in the lower atmosphere $\Gamma_1 \equiv -\partial T/\partial z$ at $z < z_1$ (see Appendix).
In Figs.~\ref{fig1}B, \ref{fig1}C, and \ref{fig1}D values of $z_1$ and a moist adiabatic $H_P$ at $\zeta = 0$ are given for $R_H$ of 80, 60 and 40\%,
respectively, as dependent on surface temperature for three representative values of $\Gamma_1$: 5, 6.5 and 9.8 K~km$^{-1}$.

We find that the existence of the undersaturated layer $z < z_1$ does not significantly change $H_P$ as compared to the case of $R_H = 100\%$.
For example, at $T_s = 300$~K we have $z_1 = 0$~km and $H_P = 5.3$~km for $R_H = 100\%$, Fig.~\ref{fig1}A, while for $R_H = 80\%$ we have $z_1 = 1.2$~km,
$H_P = 5.6$~km at $\Gamma_1 = 5.0$~K~km$^{-1}$ and $z_1 = 0.46$~km, $H_P = 5.0$~km at $\Gamma_1 = 9.8$~K~km$^{-1}$, Fig.~\ref{fig1}B. The value of $z_1$
grows with decreasing relative humidity. At $R_H = 40\%$, $T_s = 300$~K and $\Gamma_1 = 5$~K~km$^{-1}$ we have $z_1 = 4.4$~km and $H_P = 7.3$~km,
Fig.~\ref{fig1}D. However, under conditions of low relative humidity, the temperature lapse rate cannot be much smaller than the dry adiabatic lapse rate,
so the estimate of $H_P = 7.3$~km is not realistic.
In the realistic case of $\Gamma_1 = 9.8$~K~km$^{-1}$ at  $R_H = 40\%$ we have $z_1 = 1.8$~km and $H_P = 4.8$~km.
In other words, despite the condensation zone being elevated by $z_1$
when compared to the case of 100\% relative humidity, the mean height of condensation $H_P$ does not rise by the same magnitude. This is caused
by the temperature dependence of $H_P$: at $z_1 \sim ~ 1$~km, temperature $T_1$ at $z_1$ is several degrees lower than at the surface,
$T_1 < T_s$, such that the zone of intense condensation is compressed into a smaller vertical space than it would be at $T_1 = T_s$.
So the resulting $H_P$ can be even lower than in the case of $z_1 = 0$.

It is now possible to evaluate $W_t$ from (\ref{D}) at stated values of $H_P$ and $P_s$ to see if our assumption $W_t \gg |w|$ is realistic.
Equating (\ref{D}) and (\ref{Dg}) we obtain $\overline{W}_t = P_s H_P/C$, where $C \equiv \int_0^{\infty} \rho_c dz$ is the amount
of condensate and $\overline{W}_t$ is the mean vertical velocity of condensate in the column. We take the mean precipitation in the
tropical region between 30$^{\rm o}$ S and 30$^{\rm o}$ N to be $P_s = 1.3$~m~year$^{-1}$ according to the data of \citet{lw90},
a conservative (i.e. low) value of $H_P = 3.8$~km (this corresponds to $T_s = 300$~K, $R_H = 80\%$ and $\Gamma = 6.5$~K~km$^{-1}$ at $z > 0$,
curve 7 in Fig.~1B) and $C \approx 10^{-1}$~kg~m$^{-2}$ \citep{wo02}. Using these values we
obtain $\overline{W}_t \approx 1.6$ m s$^{-1}$. This value is about two orders of magnitude larger than typical time-averaged
large-scale vertical air velocities $w < 1$~cm~s$^{-1}$ observed in the tropics \citep[e.g.,][]{rex}.
We conclude that our assumption $W_t \gg |w|$ is reasonable.

We now consider the case of small terminal velocities, for which Eq.~(\ref{Dg}) does not hold. Terminal velocity depends on the size of hydrometeors and turns to zero when the condensate particles become vanishingly small.
The limiting case is the so-called ``reversible adiabat'', which corresponds to $W_t = 0$ when all condensate travels together with the air and fully evaporates in the region where the moist air descends. Note that surface precipitation in this case is zero so Eq.~(\ref{Dg}) is inapplicable.

In order to travel with the air and to reach heights that are significantly larger than $H_P$, the condensate must have a vertical velocity comparable to that of air, $w_c \simeq w$ and $W_t \ll |w|$. For $C \sim 10^{-1}$~kg~m$^{-2}$ and $W_t \ll |w| < 10^{-2}$~m~s$^{-1}$ in the tropics we obtain from (\ref{D}) $D < 10^{-2}$~W~m$^{-2}$. In other words, even if all condensate in the tropical atmosphere consisted of the smallest condensate particles, their contribution to dissipation rate would not have exceeded $10^{-2}$~W~m$^{-2}$. This means that in the tropics the small amount of condensed water that is brought by air updrafts to large altitudes significantly exceeding $H_P$ (\ref{HP}) makes little contribution to dissipation rate $D$ (\ref{Dg}), (\ref{HP}), the latter being of the order of 1~W~m$^{-2}$.

\section{Condensation incompleteness and precipitation efficiency}
\label{sec4}

As saturated moist air rises it can mix with drier air from the surroundings.  This means that its $\gamma$ (\ref{S}) drops not because of condensation,
but because of dilution through turbulent mixing: water vapor is replaced by dry air in the updraft.
In such cases condensation is incomplete: the water vapor removed from the ascending air by turbulence has not condensed. This will affect the value of the so-called
precipitation efficiency $\varepsilon$.  In empirical studies this measure, $\varepsilon$, is defined as the ratio of precipitation $P_s$ at the ground to the
inflow of moisture into the updraft. It is commonly observed to be in the vicinity of 20-40\% \citep{fa88}.

A common interpretation of these low $\varepsilon$ values from observation is that it reflects major re-evaporation of condensed water
in downdrafts \citep{ne66,fo73}. This presumes that all water vapor that has flown into the updraft will condense. Since condensation occurs
when the relative humidity is equal to unity, this logic would imply that the relative humidity within the updraft should remain high up to
the layer where the water vapor mixing ratio has dropped to a negligible value compared to its value at the surface.
For example, for a moist adiabat at $T_s = 300$~K a hundredfold reduction corresponds to a height of about 14~km, Fig.~\ref{fig2}A.
In reality, however, relative humidity drops abruptly much earlier -- for example, in hurricanes and their ambient environment it decreases sharply from over 80\% to 50-60\% at a height of about 4-5 km \citep{she69,lo90}.
The updraft of air carrying the smallest droplets high to the troposphere can continue beyond that height,
but low relative humidity means that intense condensation cannot.

In other words, precipitation efficiencies $\varepsilon < 1$ do not necessarily imply re-evaporation of condensed moisture: rather, a low ratio of precipitation to the water vapor influx can indicate an incomplete condensation. Ignoring evaporation, from a simple mass balance consideration we have $(1 - \zeta) \rho_{vs} w_s = P_s$, where $\rho_{vs}$ is the density of water vapor and $w_s$ is the vertical air velocity at the cloud base, $\zeta \equiv \gamma^*(z_2)/\gamma^*(z_1) \simeq q^*(z_2)/q^*(z_1)$ is equal to the ratio of the water vapor mixing ratio at height $z_2$ where condensation discontinues and its value at the cloud base $z = z_1$.
The flux of water vapor flowing into the cloud is given by $\rho_{vs} w_s$. In this case precipitation efficiency $\varepsilon$ is equal to $\varepsilon = P_s/(\rho_{vs} w_s) = 1 - \zeta$. If $\varepsilon = 1/3$, this means $\zeta = 2/3$, i.e. the condensation zone reaches upward to a height where the water vapor mixing ratio decreases by one third as compared to its value at the cloud base. As we show in Fig.~\ref{fig1}A, high values of $\zeta$ and, hence,
low precipitation efficiencies $\varepsilon$ are associated with relatively low precipitation path length $H_P$.

\section{Comparison with the results of \citet{pa00}}
\label{sec5}

\citet{pa00} based their estimate of $D$, the first of this kind in the meteorological literature \citep{re01,pa01}, on Eq.~(\ref{Dg}).
They noted that, if re-evaporation is neglected, $H_P$ is equal to the average height where
condensation occurs. This is correct, but some of the subsequent assumptions and derivations are less well justified. Equation~(5) of \citet{pa00} meant to define $H_P$ as $H_P = \int_0^\infty q^* dz$, where $q$ is water vapor mixing ratio, misses the normalization factor $q_s \equiv q(0)$ in the denominator (cf. our Eq.~(\ref{HP}) above at $z_1 = 0$, $z_2 = \infty$ and $z_2 \gamma^*(z_2)=0$).
Turning to quantitative estimates, \citet{pa00} proposed that the scale height of the saturated water vapor mixing ratio in the tropics is about 2.5-3~km.
\citet{pa00} do not offer any clear reasoning for this figure but refer to \citet{em96}. Our reading of \citet[p.~3284]{em96}
finds only a mention of a certain ``scale height for water vapor'' of around 3~km. However, the scale height of saturated water vapor and the scale
height of its mixing ratio are different atmospheric characteristics.
They depend differently on temperature and, hence, height. In the tropical troposphere, for example, they range from
0 km to 5 km and to 10 km, for the scale height of water vapor and its mixing ratio, respectively \citep[e.g.,][Fig.~1]{mg10}.

Leading on from their initial suggestions, \citet{pa00} offered several arguments as to why $H_P$ in (\ref{Dg}) should be several times higher than
the scale height of the saturated water vapor mixing ratio and takes a value of 5-10~km. First, they noted that the real precipitation path length $H_P$
is greater than condensation height obtained from a moist adiabat  because there is an undersaturated region in the subcloud layer. As we discussed in Section 3, while real, the effect is small and does not necessarily lead to an increase in $H_P$.

Second, \citet{pa00} proposed that an increase in $H_P$ can be induced by the entrainment of the unsaturated air parcels into the region of
saturated ascent. They did not, however, specify a mechanism for this. The entrainment of dry air causes the temperature lapse rate
to rise above the moist adiabatic value, such that the temperature drops more rapidly with height. If, despite the dry air entrainment,
the condensation nevertheless continues in the updraft prompted by this additional cooling in the ambient environment,
the change in local lapse rate will reduce rather than raise the mean condensation height $H_P$. Curve 7 in Fig.~1B illustrates
the dependence of $H_P$ on $T_s$ for $R_H = 80\%$ and a mean tropospheric lapse rate of 6.5~K~km$^{-1}$ instead of moist adiabatic lapse rate.
In this case $H_P$ depends little on surface temperature and ranges between 3 and 4~km. If, on the other hand, condensation
is discontinued by the drop in relative humidity associated with the removal of water vapor and its replacement by dry air in the region of ascent,
then $H_P$ is limited by the height where the dry air entrainment occurred. In neither case does $H_P$ increase.

\citet{pa00} also mention that the real precipitation path length is increased by the fact that some hydrometeors are lifted by updrafts to high altitudes. However, as discussed Section \ref{sec2}, such hydrometeors are small and possess terminal velocities that are much smaller than air velocity, such that Eq.~(\ref{Dg}), which \citet{pa00} intended to use, is thus inapplicable. On average, these hydrometeors make only a negligible contribution to the total dissipation rate associated with precipitation owing to both their slow terminal velocity and their small combined mass.

The final argument put forward by \citet{pa00}, and the only quantitative one, concerns re-evaporation. They presume that this effect can lead to a significant underestimate of the real value of $H_P$. \citet{pa00} quote the work of \citet{fa88} and \citet{fe96} to support the statement that a significant part (from half to two-thirds) of all condensed moisture actually re-evaporates and does not hit the ground. From this \citet{pa00} suggest that if evaporation occurs uniformly as the hydrometeors are falling, this process increases the effective precipitation path length by a factor of 1.5-2. We note for the record that \citet{fa88}, who investigated empirical data on the water budget of convective clouds including precipitation efficiency, does not mention any quantitative estimate of the re-evaporation of condensed moisture. \citet[p.~2105]{fe96}, on the other hand, do report the magnitude of re-evaporation as compared to total condensation within a squall, but their results come from a numerical model rather than observational evidence. In order to estimate the actual rate of evaporation of condensate in the downdrafts we would need to perform careful estimates of condensate transport within the cloud -- rather than measuring the transport of total moisture that is dominated by water vapor. Estimates of sufficient accuracy are not available. We can, in contrast, be confident that the effect of incomplete condensation associated with low precipitation efficiency is real. But, as discussed in the previous section, this will decrease rather than increase the estimate of $H_P$.

Furthermore, even if evaporation in downdrafts did constitute a significant fraction of total condensation, the suggestion of \citet{pa00} about a 1.5-2
increase in effective $H_P$ would still be incorrect. Let us first see how this conclusion was reached, because \citet{pa00} do not explain this
in any detail. The argument
of \citet{pa00} derives from Eq.~(\ref{Dg}) along with the one-dimensional continuity equation for condensate particles. If $j_P \equiv \rho_c(z) W_t$ is the downward flux of condensate at point $z$, then the continuity equation is $\partial j_P/\partial z = E$, where $E>0$ is evaporation. If, following \citet{pa00}, we assume that $E$ is constant
and that $j_P(0) = (1/3) j_P(H_P)$ (evaporation has decreased the original precipitation flux by two thirds as it traveled from $z = H_P$ to $z = 0$), then we have
$j_P(z) = j_P(0) (1+ 2z/H_P)$. This allows us to calculate $D$ from (\ref{D}) as $D = \int_0^{H_P} j_P g dz = 2 j_P(0) g H_P = 2 P_s g H_P$. Had this logic
been correct, we could conclude, as did \citet{pa00}, that
Eq.~(\ref{Dg}) indeed underestimates the real dissipation by half.

There are, however, two errors in this reasoning. The first one consists in neglecting the fact that local dissipation rate $\rho_c g W_t$ and local evaporation with respect to droplet size behave differently. Since absolute evaporation rate is proportional to droplet area, the smallest droplets evaporate most rapidly. If we have equal amounts $M$ (g) of small droplets with radius $r_1$ and large droplets with radius $r_2 > r_1$, the rate of depletion of total condensate from the evaporation of small droplets will be $r_2/r_1$ times faster than from large droplets (evaporation rate $\propto N s \propto (M/m) r^2 \propto M/r$, where $s \propto r^2$ is droplet's surface area, $N \propto M/m \propto M/r^3$ is the number of droplets, $m \propto r^3$ is droplet mass). In comparison, because of the fact that terminal velocity grows with increasing droplet radius, the contribution of the small droplets to total dissipation will be lower than that of the large droplets.
In theory, for spherical droplets with $W_t \propto r^2$, it will be lower by $(r_2/r_1)^2$ times. For example, with $r_2/r_1 = 10$, 90\% of all evaporation will come from droplets that make a 1\% contribution to total dissipation.

The second error in their reasoning is that they implicitly use $j_P \equiv \rho_c(z) W_t$ to represent the downward flux of condensate.
This neglects the important role of vertical air movements in transporting the smallest condensate particles.  Rather they should have used
$j_c = \rho_c(z) (w-W_t)$, where $w<0$ is the downward velocity of air. The $w$ term is particularly important in consideration of evaporation, because the smallest droplets with $W_t \ll -w$ are so slow that they can be only transported by the downdraft. This means that the continuity equation $\partial j_P/\partial z = E$ underlying the reasoning of \citet{pa00} is not valid. The correct equation $\partial j_c/\partial z = E$ does not offer any insights regarding the frictional dissipation of $D$ in the column because the distribution of $w$ remains unknown.

To summarize, we find no support for the claim of \citet{pa00} that precipitation path length should be several times higher
than the value of $H_P$ given by (\ref{HP}).

\section{Numerical estimate of $D$}

We have shown that $H_P$ grows with increasing incompleteness of condensation $\zeta$, increasing surface temperature and decreasing
lapse rate $\Gamma_1$ at $z < z_1$, Fig.~\ref{fig1}. Among these, the incompleteness of condensation $\zeta$
is both the least known and the most influential, Fig.~1A. It is closely linked to convection depth.
We suggest that the uncertainty of the mean $D$ values for the tropical region is largely determined by the uncertainty in $H_P$,
which is, in its turn, largely reflects uncertainty of $\zeta$. Taking $T_s = 300$~K as the mean surface temperature during precipitation in the tropical region, $\Gamma_1 = 5$~K~km$^{-1}$ and $R_H =80\%$, we obtain $H_P = 5.6$~km for complete condensation $\zeta = 0$, Fig.~\ref{fig1}B. At the lower end is the estimate of $H_P$ obtained assuming $\zeta = 2/3$ (corresponding to a low precipitation efficiency of $\varepsilon = 1/3$), which at $T_s = 300$~K, $\Gamma_1 = 5$~K~km$^{-1}$ and $R_H = 80\%$ gives $H_P = 2.5$~km.

\citet{pa00} estimated $D$ from the mean latent heat flux $Q$ instead of precipitation $P = Q/L_v$ in the tropics considering that $D/Q = gH_P/L_v$, where $L_v = 2.5\times 10^{6}$~J~kg$^{-1}$ is the heat of vaporization. Using $Q = 100$~W~m$^{-2}$ and $H_P$
values of 2.5 and 5.6~km we obtain a range of 1.0-2.2~W~m$^{-2}$ for the mean tropical value of $D$.

For the global mean temperature $T_s = 288$~K at $R_H = 80\%$ and $\Gamma_1 = 6.5$~K~km$^{-1}$ we have
$H_P = 3.6$~km for $\zeta = 0$, Fig.~\ref{fig1}B, and $H_P = 1.5$~km for $\zeta = 2/3$. Using the mean value of 2.5~km and mean global
precipitation of 1 m~year$^{-1}$ \citep{lv79} we obtain a global mean value of $D \sim 0.78$~W~m$^{-2}$ from (\ref{Dg}).
This means $4\times 10^{14}$~W for Earth as a whole and $1.2\times 10^{14}$~W for the gravitational power of precipitation on land.
(We note that the estimate of $10^{14}$~W for land was previously obtained based on Eq.~(\ref{Dg}) in a different context discussing renewable energy sources \citep[p.~6]{go82}.)

If the vertical profile of precipitation $P(z)$ is known, the value of $D$ can be estimated directly from (\ref{D}) under the assumption that $P(z) = \rho_c(z) W_t$. This was recently done by \citet{pa12}
who used satellite-derived $P(z)$ profiles from the Tropical Rainfall Measurement Mission and estimated $D$ for the tropical region between 30$^{\rm o}$ S and 30$^{\rm o}$ N to be 1.8~W~m$^{-2}$. Using this value \citet{pa12} went on to estimate $H_P$ from Eq.~(\ref{Dg}) by dividing $D$ (\ref{D}) by $P_s g$, $H_P = D/(P_sg)$. Having obtained values of 5.1~km for the ocean and 6.9~km for land, \citet{pa12} concluded that these results agree with their earlier estimates of 5-10~km for $H_P$ in the tropics \citep{pa00}. They proposed that the larger value obtained over land indicates  more intense convection than over the ocean.

The derivation of precipitation rates from the satellite radar data involves considerable uncertainties.
The estimation process involves a large number of empirical relationships between reflectivity and precipitation rates
as well as various assumptions concerning the properties of the hydrometeors like their size distribution and terminal velocity
\citep[e.g.,][]{uij,du98,bo05,pr09}.
The commonly used algorithms perform differently over land than they do over oceans \citep{pr09}.
They also perform differently on the ground surface versus at the top of clouds \citep{du98}.
The estimate of 1.8~W~m$^{-2}$ reported without any quantified assessment of the associated uncertainties ranges does not constrain $D$ any more
accurately than do theoretical estimates, although it can be noted that this estimate falls out of the 2-4~W~m$^{-2}$ established for the tropical region
by \citet{pa00}.

Moreover, our examination of the precipitation profiles $P(z)$ shown in Fig.~2 of \citet{pa12}
raises further problems. These data support neither the estimate of $D = 1.8$~W~m$^{-2}$ reported for the tropical region as a whole, nor the estimates
of $H_P$ made for land and ocean. Indeed, integrating these profiles
yields a value of 1.4~W~m$^{-2}$ for all the three profiles (the tropics as a whole, land and ocean). This 30\% discrepancy raises
further questions concerning the validity of the numerical estimates reported by \citet{pa12}.
The discrepancy illustrates the potential inaccuracy associated with using satellite radar data on precipitation. Heights $H_P$ estimated from the corrected values of $D$ using
$P_s = 2.6$~mm~day$^{-1}$ for ocean and $P_s = 2.1$~mm~day$^{-1}$ for land yields $H_P = 4.7$~km for the ocean and $H_P = 5.7$~km for land instead of, respectively, 5.1~km and 6.9~km obtained by \citet{pa12}.

Furthermore, the derivation of $H_P$ from Eq.~(\ref{Dg}) is potentially misleading, as such an estimate is sensitive to the estimate of surface precipitation. Satellite radar based assessments of precipitation are based on the reflectivity coefficient of precipitating particles. The radar is unable to accurately distinguish surface precipitation because of the high reflectivity of the surface. Note for example, that if surface precipitation is underestimated in the lowest kilometer, this may make little impact on the column-integrated $D$ (\ref{D}), but will have a considerable impact on the value of $H_P$. (We also note that precipitation rates in the upper part of the atmosphere can be, on the contrary, overestimated by the radar \citep{du98}.)
According to Fig.~2 of \citet{pa12}, precipitation in the lowest 1~km makes about a 1/5 contribution to the column-integrated value of $D$. If surface precipitation in this lowest region is underestimated by 25\% \citep{du98,bo05} and constitutes 75\% of the real value, this corresponds to a $25\%/5 = 5\%$ underestimate in total $D$. But $H_P = D/(g P_s)$ is then overestimated by a factor of $(100-5)\%/75\% = 1.3$. Applying this additional correction factor to our corrected $H_P$ estimates we obtain $H_P = 4.7/1.3=3.6$~km instead of 5.1~km for the ocean and $H_P = 5.7/1.3=4.4$~km instead of 6.9~km for land. Both values are outside the 5-10~km range of \citet{pa00}, but within the 3-5~km range estimated from our theoretical analysis. We also note that these values are close to $H_P = 3.8$~km that at 80\% surface relative humidity is characteristic of the mean tropospheric lapse rate 6.5~K~km$^{-1}$ in a saturated atmosphere at $T_s = 300$~K, see curve 7 in Fig.~\ref{fig1}B.

Tropical land include some exceptionally wet regions like the Amazon and Congo forests, where precipitation is 2-3 times
higher than over the nearby ocean \citep{m12}. But the tropics also include very dry regions such as the Sahara desert and the Australian interior.
Combining these wet and dry regions under one and the same category and calculating a single mean precipitation
profile for all tropical {\it land}, as done in Fig.~2 of \citet{pa12}, is of questionable value given the diversity of physical settings.
(Note that these concerns have less significance for a tropic-wide estimate the values of which are dominated by the oceans).
In the driest regions of the Earth where surface precipitation tends to zero, estimating $H_P$ from surface precipitation lacks any physical meaning,
as $H_P \to \infty$ at $P_s \to 0$. Thus any estimated value for $H_P$ does not carry any information about the real vertical distribution or
intensity of precipitation. We note in this context that while \citet{pa12} concluded that high values of $H_P$ over land are indicative of a
more intense convection, judging from their Fig.~3 one might also think that the region of most intense convection on Earth is the inner part
of Sahara desert. Here $H_P$, as estimated with use of the near zero value of surface precipitation, reaches beyond 10~km \citep[][Fig. 3]{pa12}.
This apparently nonsensical result illustrates the need to have a consistent theoretical basis for any analysis of empirical evidence.

As we discussed in Section \ref{sec2}, in a saturated atmosphere the vertical profile of precipitation $P(z)/P_s$ should coincide with the
vertical profile of $\gamma^*(z)/\gamma(0)$. In Figs.~\ref{fig2}B, \ref{fig2}C, and \ref{fig2}D the mean vertical profile of tropical precipitation
taken from Fig.~2 of \citet{pa12} is contrasted against theoretical profiles of $\gamma^*(z)/\gamma(0)$ calculated for different values of
surface temperature, relative humidity and temperature lapse rate. We see that saturated moist adiabatic profiles of $\gamma^*(z)/\gamma(0)$
tend to overestimate $P(z)/P_s$ in the upper atmosphere, Figs.~\ref{fig2}B and  \ref{fig2}C. The observed mean profile $P(z)/P_s$ is confined
between $\gamma^*(z)/\gamma(0)$ profiles built for surface relative humidity from 80\% to 40\% and having a constant lapse rate
of $\Gamma = 6.5$~K~km$^{-1}$, Fig.~\ref{fig2}D.

\section{Discussion}

In order to investigate how an atmospheric phenomenon responds to changes in atmospheric parameters it is important to establish a
sound theoretical basis concerning the key physical relationships. In this paper, building from basic physical principles and relationships,
we evaluated the rate of turbulent frictional dissipation associated with precipitation. We discussed how precipitation path length, $H_P$, is the key parameter controlling this rate, and investigated how it depends on surface temperature, humidity and the vertical extent of the area where condensation occurs.

We now consider how the frictional dissipation relates to the dynamic power of atmospheric circulation. We can illustrate this
relationship with a simple example. Consider a hanging weight tied with a short rope to an extended spring.  The system is in a state of equilibrium with the weight exactly balanced by the tension of the spring. When we cut the rope the weight can fall. At this moment two types of potential energy have become available: the first is the weight's energy in the gravitational field and the second is the energy of the stretched spring.
The first potential energy is transformed to kinetic energy as the weight falls, while the second potential energy is transformed to kinetic energy as the spring accelerates upward. It is clear from this example that the two potential energies are independent in magnitude: the first depends on the initial height of the weight above the surface, the second depends on the elasticity of the spring and is independent of height and of what happens to the weight after the rope is cut.

Likewise upon condensation, two types of potential energy are formed: first the potential energy of falling hydrometeors as dictated by the precipitation path length $H_P$ (\ref{HP}) and second the potential energy of the non-equilibrium pressure gradient that results from the disappearance of water vapor from the gas phase. The key peculiarity of this second process consists in the fact that this potential energy is coupled to the vertical motion of moist air and is released only when the air moves upwards.

Since precipitation and condensation always accompany each other, it is of interest to compare the powers associated with each of the two processes
and how these might influence atmospheric motion.
Previously we have argued that the dynamic power of condensation-induced circulation $D_c$ (W~m$^{-2}$) per unit surface area can be estimated
as $D_c = P_sRT/M_v$,  where $R = 8.3$~J~mol$^{-1}$~K$^{-1}$ is the universal gas constant and $T$ is the mean temperature in the atmospheric column where condensation occurs
\citep{mg10,mg11,g12}. This formula results from the proposition that the dynamic power ${\cal D}_c$ (W~m$^{-3}$)  per unit air volume of the upward pressure gradient force induced by condensation and associated with the non-equilibrium vertical gradient of saturated water vapor, is equal to
\begin{eqnarray}\label{DDc}
{\cal D}_c = w\Big(\frac{\partial p_v^*}{\partial z} - \gamma \frac{\partial p}{\partial z}\Big) = wp \frac{\partial \gamma^*}{\partial z}
 = RT S,
\end{eqnarray}
where $S = w N \partial \gamma^*/\partial z$ is the net condensation rate per unit volume in mol~H$_2$O~m$^{-3}$~s$^{-1}$, $N$ (mol~m$^{-3}$) is molar
density of air, see (\ref{S}). Neglecting the minor dependence of $T \simeq T_s$ on $z$ and observing that the integral of $S$ over $z$ is equal to $P_s/M_v$, see Appendix, we have
\begin{equation}\label{Dc}
D_c = \int_0^\infty {\cal D}_c dz = \frac{P_sRT_s}{M_v}.
\end{equation}

To illustrate the value of this theoretical result we can now estimate the global power of condensation-induced circulation. Using a global mean value of $P_s$ of 1 m~year$^{-1}$ \citep{lv79,lw90} and global mean surface temperature of $T_s = 288$~K, we have $D_c = 4$~W~m$^{-2}$. For the tropical region with mean $P_s = 1.3$~m~year$^{-1}$ we have $D_c = 5.5$~W~m$^{-2}$. Since under conditions of hydrostatic equilibrium the work done by the vertical pressure gradient is compensated by the work done by gravity, the kinetic energy of the large-scale air flow derives from the horizontal pressure gradient alone. The power of this horizontal force per unit air volume is equal to $u \partial p/\partial x$, where $u$ is the horizontal velocity component parallel to the horizontal pressure gradient. The dynamic power of atmospheric circulation (the rate at which the kinetic energy is generated) can therefore be estimated from the observed horizontal pressure gradients and the observed $u$ values. It can also be estimated as the power of turbulent dissipation of the air flow under the assumption
that in the stationary case the dynamic power that creates the kinetic energy is equal to the power of turbulent dissipation $D_t$ of this energy. The available global mean observation-based estimates of these powers are in the range of 2-4 W~m$^{-2}$ \citep{oo64,lo67,pe92}. Our global mean estimate, derived from basic principles, falls at the upper edge of this range.

The ratio of the powers of the two processes - the frictional dissipation power $D$ of hydrometeors and the dynamic power $D_c$ of condensation-induced circulation - is given by
\begin{equation}
\label{rat}
\frac{D}{D_c} = \frac{H_P}{h_v}, \quad h_v \equiv \frac{RT_s}{M_v g}.
\end{equation}
This ratio does not depend on precipitation rate $P_s$ but grows with temperature owing to the dependence of $H_P$ on $T_s$, Fig.~\ref{fig1}.
The value of $h_v$ (\ref{rat}) has the meaning of the scale height of (unsaturated) water vapor in hydrostatic equilibrium, at $T_s = 300$~K it is
equal to 14~km. With the large-scale values of $H_P$ not exceeding 6 km at mean surface temperatures not exceeding 300~K, Fig.~\ref{fig1}, we obtain a
general estimate of $D/D_c < 0.4$.

Noting that $D$ grows with surface temperature, Fig.~1, we can estimate when $D$ exceeds $D_c$. We do this using (\ref{rat}) and
assuming that the budget of energy turnover for a condensation-induced circulation has the form of $D_c = D_t + D$, where $D_t$ is turbulent dissipation of the large-scale air flow.
At $R_H = 80\%$ and $\Gamma_1 = 5$~K~km$^{-1}$ $D$ equals $D_c$ at $T_s = 323$~K, Fig.~\ref{fig1}B.
At  higher temperatures in a moist adiabatic atmosphere  any significant circulation due to condensation will be prevented because of the
insufficient dynamic power to overcome the energy losses associated with frictional dissipation due to precipitation.
If the atmosphere is not moist adiabatic but has a constant lapse rate of $\Gamma = 6.5$~K~km$^{-1}$, $D$ grows much more slowly
with increasing surface temperature (see Fig. 1, curves 7). It does not approach $D_c$ anywhere at $T_s < 360$~K, i.e. in the entire range
where the approximation $\gamma \ll 1$ on which Eq.~(\ref{HP}) is based holds.
From these considerations we conclude that frictional dissipation due to precipitation is insufficient to arrest
condensation-induced atmospheric circulation on Earth.

\ifthenelse{\boolean{dc}}
{}
{\clearpage}
\begin{appendix}
\section*{
\begin{center}
Equations for calculating $z_1$ and $H_P$
\end{center}}

Eq.~(\ref{HP}) is obtained using (\ref{S}) and considering that
\begin{eqnarray}
\int\limits_{z_1}^{z_2} S(z) z  dz  & =&  w N \gamma^* z \Big|_{z_2}^{z_1} + \int_{z_1}^{z_2} w N \gamma^* dz
+ \int_{z_1}^{z_2} \frac{\partial (wN)}{\partial z }\gamma^* z dz  \nonumber\\
&\simeq& w N \left\{\gamma^* z \Big|_{z_2}^{z_1} + \int_{z_1}^{z_2} \gamma^* dz\right\} , \label{parts1}\\
\int_{z_1}^{z_2} S(z) dz  & =&  w N \gamma^* \Big|_{z_2}^{z_1}
+ \int_{z_1}^{z_2} \frac{\partial (wN)}{\partial z }\gamma^* dz \nonumber\\
&\simeq&  w N \gamma^* \Big|_{z_2}^{z_1} .   \label{parts2}
\end{eqnarray}
From the one-dimensional stationary continuity equation we have $\partial (wN)/\partial z = -S$. This means
that the terms discarded in  \eqref{parts1} and \eqref{parts2} constitute a small magnitude of the order of $\gamma^* \ll 1$ as compared to the initial terms. Quantity $wN$ changes little with $z$ (by a relative magnitude of the order of $\gamma^*$) as compared to $\gamma^*$ that changes severalfold. Therefore $Nw$ can be assumed to be constant and cancelled from both the denominator and nominator in ratio (\ref{HP}). The inaccuracy of the resulting expression for $H_P$ (\ref{HP}) is of the order of $\gamma^*$ and, for temperatures of interest, does not exceed 10\%.

\begin{figure*}[h]
\begin{minipage}[h]{0.49\textwidth}
\centerline{
\includegraphics[width=0.99\textwidth,angle=0,clip]{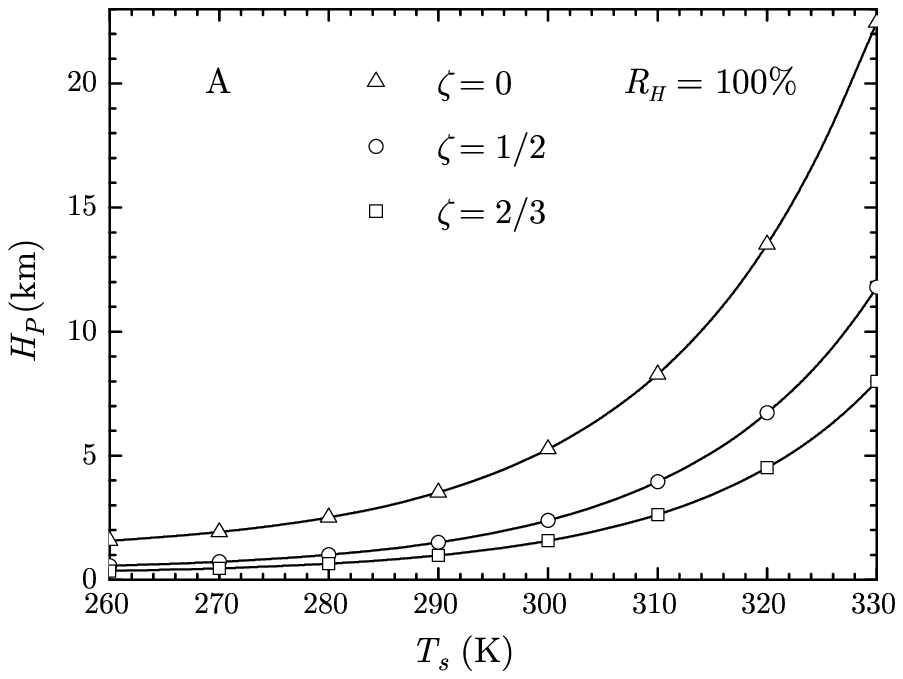}
}
\end{minipage}
\begin{minipage}[h]{0.49\textwidth}
\centerline{
\includegraphics[width=0.99\textwidth,angle=0,clip]{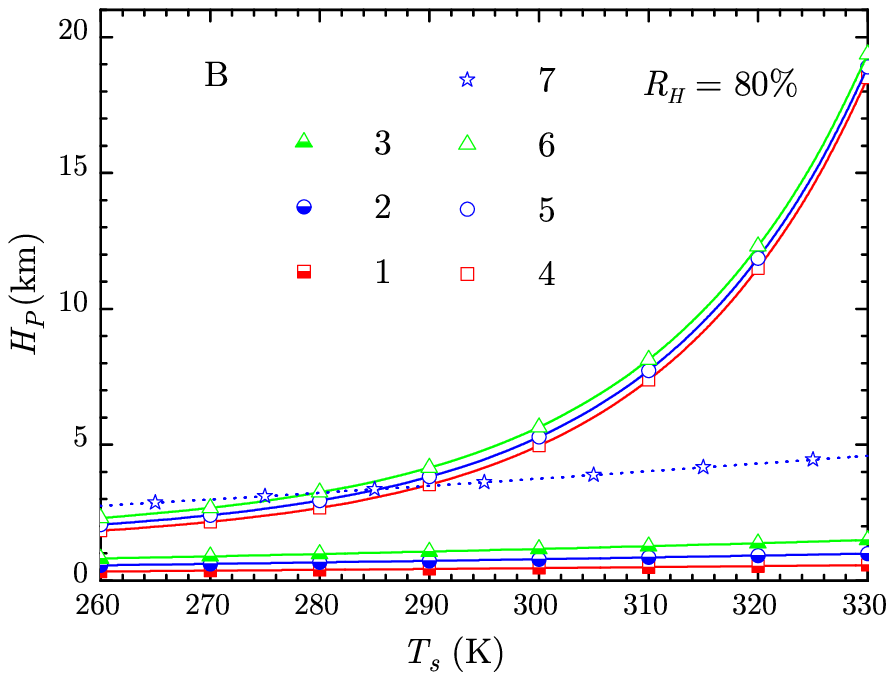}
}
\end{minipage}
\begin{minipage}[h]{0.49\textwidth}
\centerline{
\includegraphics[width=0.99\textwidth,angle=0,clip]{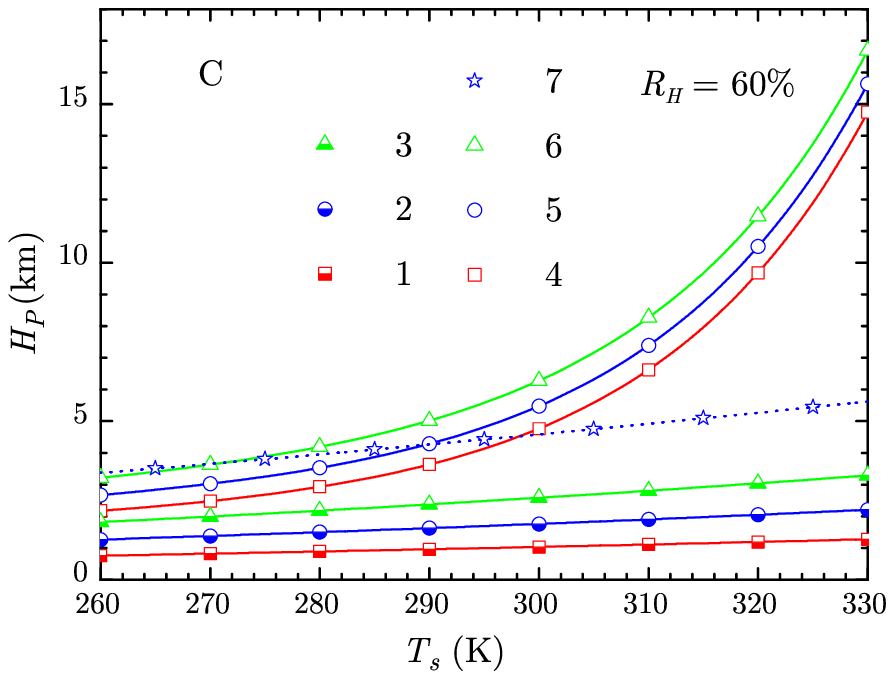}
}
\end{minipage}
\begin{minipage}[h]{0.49\textwidth}
\centerline{
\includegraphics[width=0.99\textwidth,angle=0,clip]{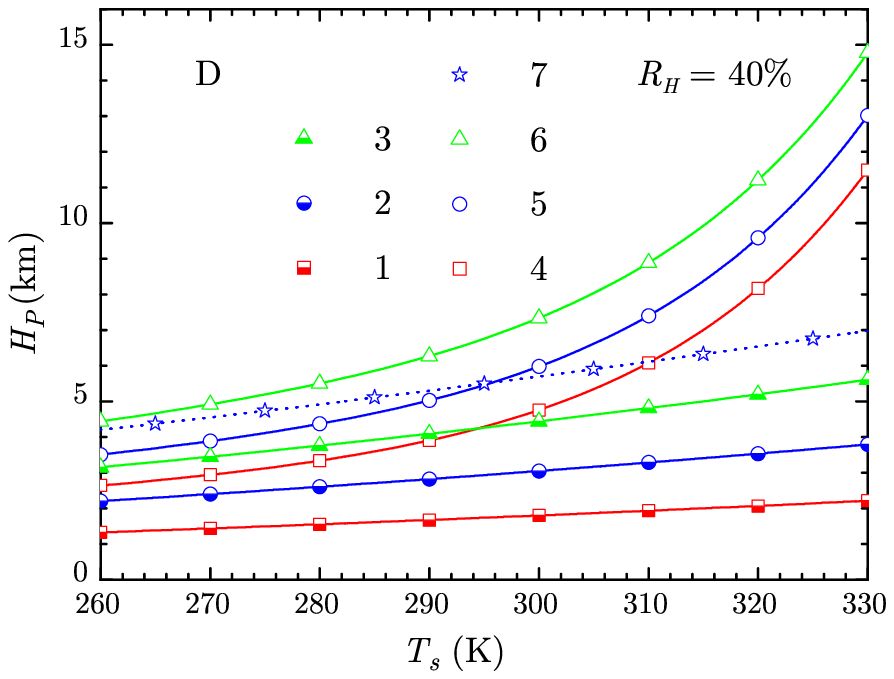}
}
\end{minipage}
\caption{\label{fig1}
Precipitation path length $H_P$ as a function of surface temperature $T_s$ at various values of
surface relative humidity $R_H$: 100\% (A), 80\% (B), 60\% (C), 40\% (D).
\newline
A: Moist adiabat in a fully saturated atmosphere, $z_1 = 0$. Height of condensation area $z_2$
is equal to the height where $\zeta \equiv \gamma^*(z_2)/\gamma^*(z_1) = 0$ (triangles), $\zeta = 1/2$ (circles) or $\zeta = 2/3$ (squares).
\newline
B, C, D: Curves 1 -- 3 denote $z_1$, curves 4 -- 7 denote $H_P$ at $\zeta=0$. Temperature lapse rate $\Gamma_1$ at $z < z_1$ is equal to 9.8~K~km$^{-1}$ (red squares), 6.5~K~km$^{-1}$ (blue circles, stars), 5~K~km$^{-1}$ (green triangles). At $z \geqslant z_1$ the atmosphere is moist adiabatic, except for curves 7, which are for an atmosphere that is saturated above $z_1$ and has a constant temperature lapse rate $\Gamma =6.5$~K~km$^{-1}$ everywhere at $z \geqslant 0$ (see Appendix for details).
}
\end{figure*}

\begin{figure*}[h]
\begin{minipage}[h]{0.49\textwidth}
\centerline{
\includegraphics[width=0.99\textwidth,angle=0,clip]{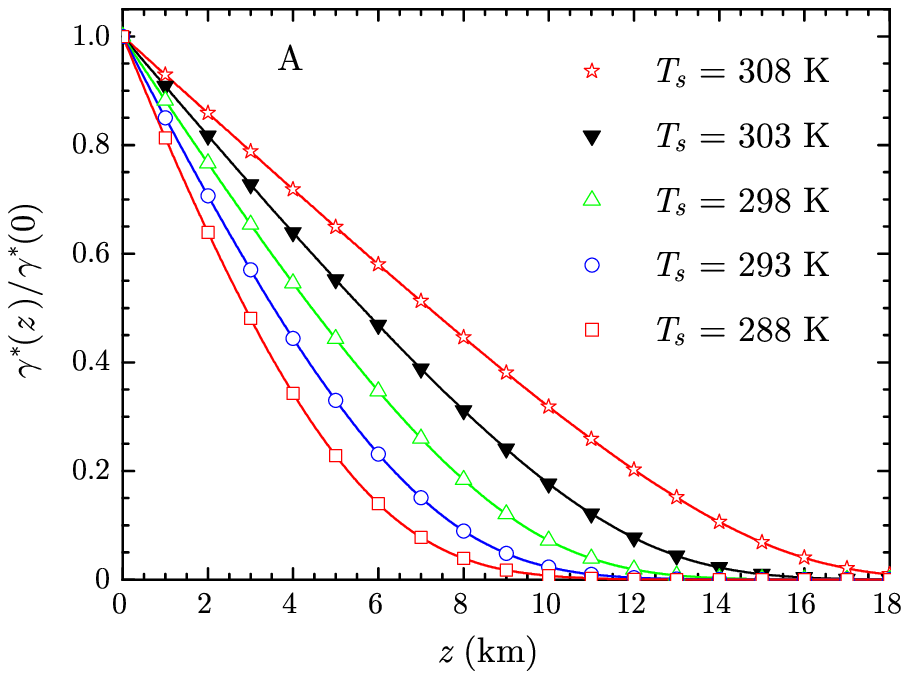}
}
\end{minipage}
\begin{minipage}[h]{0.49\textwidth}
\centerline{
\includegraphics[width=0.99\textwidth,angle=0,clip]{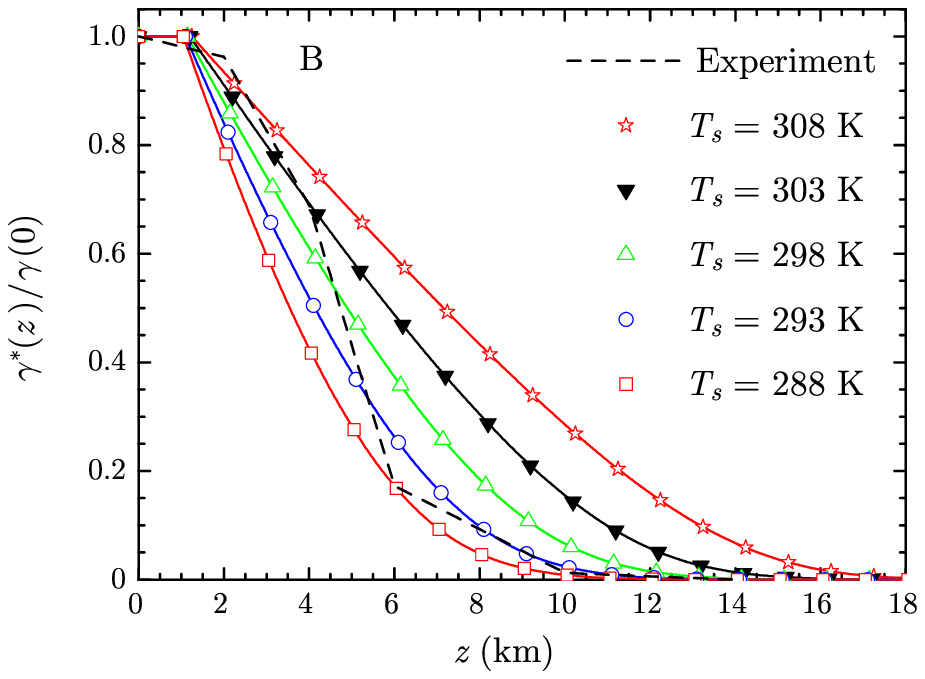}
}
\end{minipage}
\begin{minipage}[h]{0.49\textwidth}
\centerline{
\includegraphics[width=0.99\textwidth,angle=0,clip]{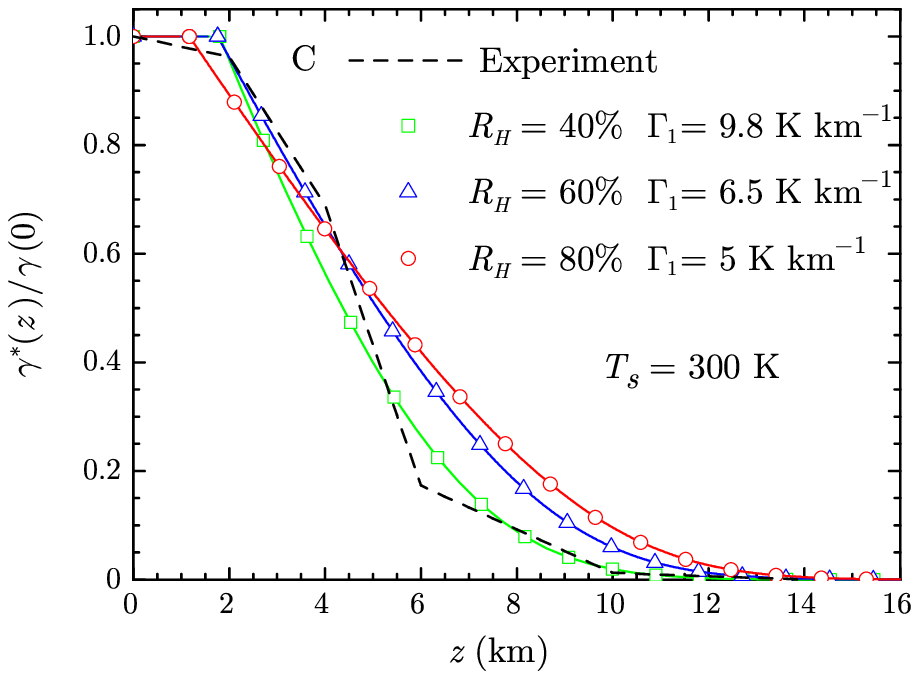}
}
\end{minipage}
\begin{minipage}[h]{0.49\textwidth}
\centerline{
\includegraphics[width=0.99\textwidth,angle=0,clip]{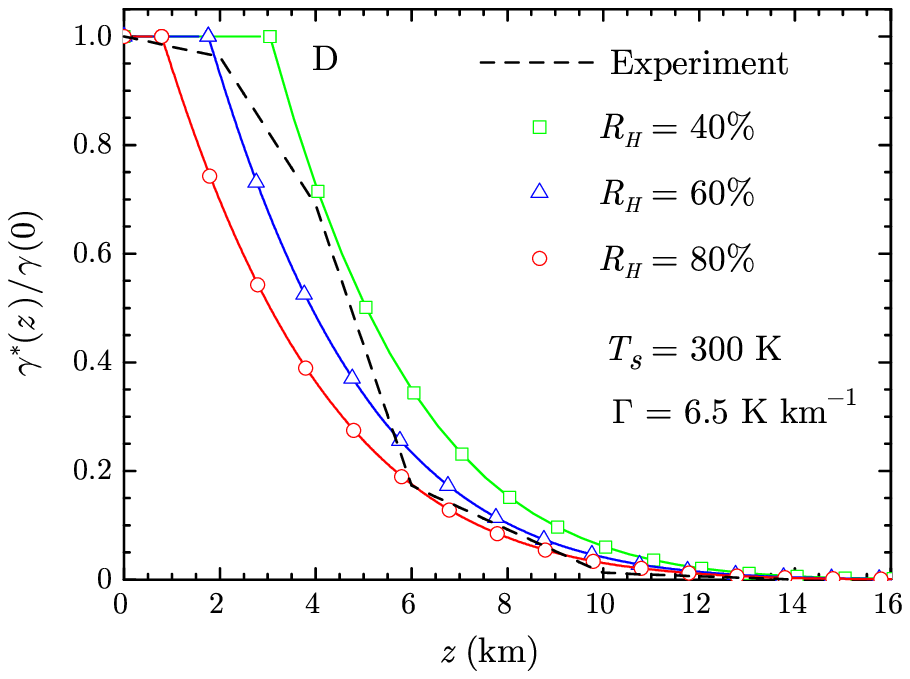}
}
\end{minipage}
\caption{\label{fig2}
Vertical distribution of $\gamma^*(z)/\gamma(0)$ under different atmospheric conditions. Dashed curve denotes the mean observed tropical precipitation $P(z)/P(0)$ from Fig.~2 of \citet{pa12}.
\newline
A: Moist adiabat in a completely saturated atmosphere, $R_H = 100\%$, cf. Fig.~1A.
\newline
B: The atmosphere is unsaturated at $z < z_1$ ($R_H = 80\%$, $\Gamma_1 = 5$~K~km$^{-1}$) and moist adiabatic at $z \geqslant z_1$, cf. Fig.~\ref{fig1}B.
\newline
C: The atmosphere is unsaturated at $z < z_1$ and moist adiabatic at $z \geqslant z_1$.
\newline
D: The atmosphere is unsaturated at $z < z_1$, saturated at $z \geqslant z_1$ and has a constant temperature lapse rate $\Gamma = 6.5$~K~km$^{-1}$ at $z \geqslant 0$ (cf. curves 7 in Figs.~\ref{fig1}B, \ref{fig1}C, and \ref{fig1}D).
}
\end{figure*}

The system of equations for moist adiabat solved to calculate $H_P$ in Fig.~\ref{fig1}A is as follows \citep{mg10,g12}:
\begin{gather}
\frac{1}{T}\frac{\partial T}{\partial z} - \frac{\mu}{p} \frac{\partial p}{\partial z}+ \frac{\mu \xi}{1-\gamma^*} \frac{\partial \gamma^*}{\partial z} = 0,  \label{eqA1}\\
\frac{1}{\gamma^*}\frac{\partial\gamma^*}{\partial z} -  \frac{\xi}{T}\frac{\partial T}{\partial z}  + \frac{1}{p}
\frac{\partial p}{\partial z} =0, \label{eqA2}\\
-\frac{1}{p} \frac{\partial p}{\partial z} - \frac{M g}{RT}=0,  \label{eqA3}
\end{gather}
where temperature $T$, air pressure $p$, and $\gamma^* \equiv p_v^*/p$ -- the relative partial pressure of saturated water vapor -- are functions of height $z$. Here $\xi \equiv L_v/(RT)$, $L_v = 45 \times 10^{3}$~J~mol$^{-1}$ is heat of vaporization, $R = 8.3$~J~mol$^{-1}$~K$^{-1}$ is the universal gas constant, $M = (1-\gamma^*)M_d + \gamma^* M_v$, $M_d=29$ g~mol~$^{-1}$,  $M_v=18$ g~mol$^{-1}$, $\mu = R/c_p= 2/7$, $c_p$ is the molar heat capacity of air at constant pressure (J mol$^{-1}$ K$^{-1}$). Equation \eqref{eqA1} results from the first law of thermodynamics for moist air saturated with water vapor. Equation \eqref{eqA2} derives from the definition of $\gamma^*$ combined with the Clausius-Clapeyron law. Equation \eqref{eqA3} is equivalent to the condition of hydrostatic equilibrium \eqref{he} for ideal gas.

The boundary conditions for the surface $z=0$ at a given surface temperature $T_s$ read
\begin{gather}
T=T_s,  \label{eqA4}\\
p=p_s, \label{eqA5}\\
p^*_v(T)= p^*_{v0} \exp(\xi_0 - \xi),   \label{eqA6}
\end{gather}
where $p^*_{v0}$ and $\xi_0 = L_v/(RT_0)$ correspond to some reference temperature $T_0$. We take $T_0=303$ K, $p^*_{v0} = 42$ hPa and the standard value
for the atmospheric pressure $p_s=1013$ hPa. The dependence of vaporization heat $L_v$ on temperature is neglected. In this case $\xi_0 = 18$.

For a fully saturated atmosphere $z_1=0$ in (\ref{HP}). Numerical evaluation of the system of Eqs.~\eqref{eqA1}-\eqref{eqA3} allows us to obtain the unknown functions $T(z)$, $p(z)$, $\gamma^*(z)$ and to calculate $H_P$ \eqref{HP} as shown in Fig.~\ref{fig1}A.

In Figs.~\ref{fig1}B, \ref{fig1}C, and \ref{fig1}D the atmosphere at the surface $z = 0$ is not saturated and has a relative humidity of 80\%, 60\% and 40\%, respectively. To find height $z_1$ where the relative humidity reaches unity, we assume that within the range $0 \leqslant z \leqslant z_1$ the non-saturated $\gamma =p_v/p$ is constant and temperature $T(z)$ drops versus height $z$ at a constant lapse rate $\Gamma_1$, $T(z) = T_s -  \Gamma_1 z$. Then the non-saturated pressure $p_v$  of water vapor is given by
\begin{equation}\label{eqA7}
p_v(z)=  p_v^*(T_s)R_H \frac{p(z)}{p_s},
\end{equation}
where $R_H$ is relative humidity at the surface $z=0$, saturated pressure $p_v^*$ of water vapor is governed by
the Clausius-Clapeyron law \eqref{eqA6} and pressure $p(z)$ conforms to the condition of hydrostatic equilibrium \eqref{eqA3} with $M \simeq M_d$. Height $z_1$ where relative humidity becomes unity and condensation commences, is a function of the surface temperature $T_s$. We find height $z_1$ as the solution of equation
\begin{equation}
\label{z1}
p_v(z_1)= p_v^*(T_1), \quad T_1 \equiv T_s - \Gamma_1 z_1.
\end{equation}

The atmosphere is assumed to be saturated and moist adiabatic within the range $z_1 \leqslant z \leqslant z_2$. We need to evaluate the system of Eqs.~\eqref{eqA1}-\eqref{eqA3} in order to find functions $T(z)$, $p(z)$, and $\gamma^*(z)$. However, the boundary conditions should now be imposed not at the ground surface but at $z=z_1$, so that $T=T_1$ and $p=p(z_1)$ instead of $T_s$ \eqref{eqA4} and $p_s$ \eqref{eqA5} at $z=0$. Using the obtained solutions, condensation rate is calculated from (\ref{HP}).

Curves 7 in Figs.~\ref{fig1}B, \ref{fig1}C, and \ref{fig1}D are obtained for an atmosphere that is unsaturated at $z < z_1$, saturated
at $z \geqslant z_1$ and has a constant temperature lapse rate of $\Gamma = 6.5$~K~km$^{-1}$ at $z \geqslant 0$.
The solution at $z \geqslant z_1$ is obtained by solving the system of Eqs.~(\ref{eqA2}), (\ref{eqA3}) and $\partial T/\partial z = -\Gamma$ instead of (\ref{eqA1}).

\end{appendix}

\ifthenelse{\boolean{dc}}
{}
{\clearpage}
\bibliographystyle{ametsoc}
\bibliography{fric-ref}


\end{document}